\documentstyle[manuscript,aps]{revtex}
\begin{document}

\title{Quantum M\"{u}nchhausen effect in tunneling}

\author{V.V. Flambaum$^{1}$ and 
V.G. Zelevinsky$^{2}$\thanks{email address: zelevinsky@nscl.msu.edu}}

\address{$^1$ School of Physics, University of New South Wales,
Sydney 2052, Australia}

\address{$^2$ Department of Physics and Astronomy and National Superconducting
Cyclotron Laboratory, Michigan State University,
 East Lansing, Michigan 48824, USA}

%\twocolumn[
%\date{\today}
\maketitle
%\widetext

%\vspace*{-1.0truecm}

\begin{abstract}
%\begin{center}
%\parbox{14cm}
It is demonstrated that radiative corrections increase tunneling
probability of a charged particle.
%\end{center}
\end{abstract}

\pacs{PACS numbers:  03.65.Sq, 12.20-m, 24.90.+d}

%] \narrowtext

Famous baron von M\"{u}nchhausen saved himself from
a swamp pulling his hairs by the hands of his own \cite{Mun}. 
According to classical physics, such a feat 
seems to be impossible. However, we live in a quantum world.
In a tunneling of a charged particle, the head of the particle
wave function can send a photon to the tail which absorbs this photon
and penetrates the barrier with enhanced probability. 
Obviously, such a photon feedback should work
in the two-body tunneling where the first particle, while continuing
to be accelerated by the potential after the tunneling,
can emit a (virtual) photon that increase energy of the second particle
and its tunneling probability. The M\"{u}nchhausen mechanism may be helpful
in the tunneling of a composite system. It is related to phonon assisted
tunneling but does not require any special device being always provided by the
interaction of a charged particle with the radiation field. 

The interaction of
a tunneling object with other degrees of freedom of the system and the
influence of this interaction on the tunneling probability for a long time was
a topic of intensive studies initiated by Caldeira and Leggett \cite{cal}.
Their general conclusion, in agreement with intuitive arguments, was that any
friction-type interaction suppresses the tunneling. At the same time, it was
realized that such an interaction leads to distortions of the barrier which can
be helpful in endorsing the tunneling. The simplest effect is associated with
the zero-point vibrations of the source responsible for the existence of the
barrier. This is important for the probabilities of subbarrier nuclear
reactions as pointed out by Esbensen \cite{esb}. In the last decade, many
experimental and theoretical efforts were devoted to the understanding of
related aspects of subbarrier reactions, see the recent review \cite{bal}
and references therein. Below we discuss the interaction of a charged
tunneling object with the electromagnetic field which always accompanies motion
of the object.

Formally speaking, we are looking for the effects of radiative corrections on 
the single-particle tunneling. These effects can be described by
the Schr\"{o}dinger equation with the self-energy operator:
\begin{equation}
\label{H}
\hat{H}\Psi({\bf r}) + \int \Sigma({\bf r},{\bf r}';E) \Psi({\bf r}') d^3r'= 
E \Psi({\bf r})
\end{equation}
where $\hat{H}$ is the unperturbed particle hamiltonian, which includes a 
barrier potential, and $\Sigma=M-i\Gamma/2$ is the complex nonlocal and 
energy-dependent operator determined by the coupling to virtual photons 
and a possibility of a real photon emission. The ``photon hand" here connects 
two points ${\bf r}$ and ${\bf r}'$ of the same wave function. 
In the one-photon approximation the self-energy due to the interaction with 
the transverse radiation field can be written as 
\begin{equation}
\Sigma({\bf r},{\bf r}';E)=\sum_{{\bf k},\lambda}|g_{{\bf k}}|^{2}
\sum_{n}\frac{\langle {\bf r}|(\hat{{\bf p}}\cdot {\bf e}_{{\bf k}\lambda})
e^{i{\bf k\hat{r}}}
|n\rangle\langle n|(\hat{{\bf p}}\cdot{\bf e}^{\ast}_{{\bf k}\lambda})
e^{-i{\bf k\hat{r}}}|{\bf r}'\rangle}{E-E_{n}-\omega_{{\bf k}}-i0}. \label{1}
\end{equation}
Here we sum over unperturbed stationary states $|n\rangle$; $\hat{{\bf r}}$
and $\hat{{\bf p}}$ are the position and
momentum operators, respectively;
the photons are characterized by the momentum ${\bf k}$,
frequency $\omega_{{\bf k}}$ and polarization $\lambda$; the polarization
vectors ${\bf e}_{{\bf k}\lambda}$ are perpendicular to ${\bf k}$ so that the
momentum operators commute with the exponents.
The normalization factors are included into $g_{{\bf k}}\propto 
\omega_{{\bf k}}^{-1/2}$. The relativistic generalization of (\ref{1}) 
is straightforward.

The hermitian part $M$ 
of the self-energy operator is given by the principal value
integral over photon frequencies in (\ref{1}). The expectation value of $M$ is 
responsible for the Lamb shift of bound energy levels. It contains also the mass
renormalization for a free particle which should be subtracted. Our problem is
different from the energy shift calculation for bound states since we are
interested in the change of the wave function of the tunneling particle.
However we can use some features of the conventional approach.
As well known from the 
Lamb shift calculations, one can use different approximations in the
two regions of integration over the photon frequency $\omega$. 
In the nonrelativistic low-frequency 
region, $\omega<\beta m$, where the parameter $\beta<1$ is chosen in such a
way that typical excitation energies of a particle in the well $\delta E$ are
smaller than $\beta m$ (in the hydrogen Lamb shift problem a fine
structure constant $\alpha$ can play the role of the borderline scale 
parameter), it is possible to neglect 
the exponential factors in (\ref{1}). The high-frequency 
contribution to $M$, where the potential can be
considered as a perturbation to free motion, has been calculated, e. g.
 in Ref. \cite{Akhieser}. The two contributions match smoothly at
 $\omega =\beta m$. 

It is easy to estimate the mass operator $M$ with logarithmic accuracy.
After summation over
polarizations and standard regularization \cite{Akhieser}, the low frequency
part of the operator $M$ can be written as 
\begin{equation}
\label{sigma}
\hat{M}(E)=\frac{2 Z^2 \alpha}{3 \pi m^{2}} \int d\omega
\sum_n \hat{{\bf p}}|n\rangle
\frac{E - E_n}{E - E_n -\omega}\langle n|\hat{{\bf p}}
\end{equation}
where $Ze$ is the particle charge,
and $m$ is the mass of the particle (reduced mass in the alpha-decay case).
We use the units $\hbar=c=1$. 
Substituting the logarithm arising from the frequency integration by its average
value $L= \ln(\beta m/\omega_{min}) $, we can use the closure relations
 and obtain a simple expression
\begin{eqnarray}
\hat{M}(E)&=&\frac{2Z^{2}\alpha}{3\pi m^{2}}L\hat{{\bf p}}(\hat{H}
-E)\hat{{\bf p}} \\
&=&\frac{Z^{2}\alpha}{3\pi m^{2}}L\left\{\nabla^{2}\hat{U}
+[(\hat{H}-E),\hat{{\bf p}}^{2}]_{+}
\right\}.                                                  \label{2}
\end{eqnarray}
The mean value of the term with the anticommutator $[...,...]_{+}$ in eq. 
(\ref{2}) is equal to zero since $(\hat{H}-E)\Psi_0=0$ where $\Psi_0$ is the 
unperturbed wave function. A correction to the wave function due to this term 
can be calculated by using perturbation theory and the unperturbed
Schr\"{o}dinger equation,
\begin{equation}
\delta\Psi=\frac{2Z^{2}\alpha}{3\pi m}L[U-\langle 0|U|0\rangle]\Psi_{0}.
                                                               \label{2a}
\end{equation}
This correction is not essential since it does not influence the exponent in
the tunneling amplitude.

Combining the remaining term in eq. (\ref{2}) with
the high-frequency contribution which contains 
 $L= \ln(m/\beta m)$, see  Ref. \cite{Akhieser}, the result 
can be presented as an effective local 
operator proportional to the Laplacian $\nabla^{2}U({\bf r})$,
\begin{equation}
\label{sigmaloc}
M({\bf r}, {\bf r}'; E) \simeq \nabla^{2}U({\bf r}) \delta({\bf r}-
{\bf r}')\frac{ Z^2 \alpha}{3 \pi m^2}
 \ln\frac{m}{U_0} \equiv \delta U({\bf r})\delta({\bf r}-{\bf r}').
\end{equation}
Here we used the barrier height $U_0$ as a lower
cut-off $\omega_{min}$ of the integration over frequencies (below we give a 
semiclassical estimate which leads also to a more accurate evaluation of the
logarithmic factor). For the tunneling of an extended
object,
the mass $m$ in the argument of the logarithm should be replaced by the
inverse 
size of the particle $1/r_0$ which comes from the upper frequency cut-off
given in this case by the charge formfactor. 
The obtained result is
physically equivalent to the averaging over the position fluctuations due to
the coupling to virtual photons.
Thus, in the logarithmic approximation the mass operator is reduced to a local
correction  $\delta U({\bf r})$ to the potential $U({\bf r})$.

The Laplacian of the potential energy $\nabla^{2} U({\bf r})$ near 
the maximum of the barrier
is negative (correspondingly, near the bottom of the potential well 
it is positive). Therefore, we obtained the negative
correction $\delta U({\bf r})$ to the potential  barrier which leads to a 
conclusion that jiggling of the photon
increases the tunneling amplitude of the particle. The
numerical value of the correction to the potential is small, $\sim$ 1 keV, for
the alpha-decay. However, in some cases it  may be noticeable due to
the exponential dependence on the height of the barrier (recall the notorious
cold fusion problem). Also, there exist theories like QCD where the radiation
 corrections are not small. In many-body systems one can use collective 
modes, as phonons, to transfer energy. This can influence electron tunneling 
through quantum dots or insulating surfaces. Energy exchange between a
tunneling particle and nuclear environment is known to be 
important in subbarrier nuclear fission and fusion \cite{bal}.

An analysis can be performed more in detail by using the semiclassical WKB
approximation for the tunneling wave functions. The semiclassical radial
 Green function of
unperturbed motion under the barrier can be written in terms of the classical
momentum in the forbidden region, $p(r;E)=[2m(U(r)-E)]^{1/2}$, 
at a given energy $E$ as
\begin{equation}
G(r,r';E)=-m[p(r)p(r')]^{-1/2}\left\{e^{-\int_{r'}^{r}d\xi\,p(\xi)}\Theta(r-r')
+e^{-\int_{r}^{r'}d\xi\,p(\xi)}\Theta(r'-r)\right\}          \label{3}
\end{equation}
where $\Theta(x)$ is the step-function. The full three-dimensional Green
function
 $G(E)=\sum_n |n\rangle(E-E_{n})^{-1}\langle n|$
contains also angular harmonics which could be separated in a
routine way accounting for the fact that in the long wavelength approximation
for the $s$-wave solution the intermediate states are $p$-waves. Indeed, the
operator of electric dipole radiation ${\bf \hat p}$ converts an initial
 $s$-wave  $\Psi$ in eq. (\ref{H})  into
an intermediate  $p$-wave state $|n\rangle$. Therefore, it is sufficient
to keep the $p$-wave part of the radial Green function and to use closure in 
the sum over angular harmonics. 

The kernel of the integral term in the Schr\"{o}dinger equation (1) contains 
\begin{equation}
K(r,r';E)=\int d\omega\,G(r,r';E-\omega).                     \label{4}
\end{equation}
The integrand consists of terms falling exponentially as $|r-r'|$ increases.
The potential $U(r)$ is assumed to be a smooth function. Therefore, we can put
 $p(r')\approx p(r)$. Now it is easy to perform the integration over
$\omega$ in eq.(\ref{4}) which leads to
\begin{equation}
K(r,r';E)=-\frac{1}{|r-r'|}\left\{e^{-p_{min}|r-r'|} -
e^{-p_{max}|r-r'|}\right\}                     \label{4a}
\end{equation}
where $p_{min}=[2m(U_{p}(r)-E)]^{1/2}$ ,  $p_{max}=[2\beta ]^{1/2}m$, and 
$U_{p}(r)$ is the effective $p$-wave radial
potential which includes the centrifugal
part. This expression has a very narrow maximum near $r=r'$ with the width
$|r-r'| \sim 1/p_{max}$. This is a measure of non-locality of the
self-energy operator $M(r,r';E)$. In any nonrelativistic application
the kernel can be treated as proportional to the delta-function. The 
proportionality coefficent
can be found by the integration over $r$. Thus, we obtain the local behavior
 of the kernel,
\begin{equation}
K(r,r';E)\approx -L(r)\delta(r-r'),                \label{5}
\end{equation}
where now we determine the lower limit of the logarithm which has appeared in
our previous derivation (\ref{sigmaloc}) as related to the local value of the
potential,
\begin{equation}
L(r)=\ln\frac{m}{|U_{p}(r)-E|}.                                 \label{6}
\end{equation}
 The substitution into eq. (\ref{sigmaloc}) gives
\begin{equation}
\label{deltaU}
 \delta U({\bf r})= \frac{ Z^2 \alpha}{3 \pi m^2} \ln\frac{m}{|U_{p}(r)-E|}
\nabla^{2}U({\bf r}).
\end{equation}
As usual, this semiclassical expression is not valid near the
turning points where $U_{p}(r)=E$. However, a very weak logarithmic singularity
does not produce any practical limitations on the applicabilty
of eq. (\ref{deltaU}).

  The conclusion of enhancement of the tunneling probability 
seems to contradict to the common sense: radiation should
cause energy losses and reduce the tunneling amplitude of the charged
particle. However, such an argument may be valid only for the 
real photon emission.
This emission is described by the antihermitian part of the self-energy 
operator which is originated from the delta-function corresponding to on-shell
processes,
\begin{equation}
\label{gamma}
\Gamma({\bf r},{\bf r}';E) =\frac{4 Z^2 \alpha}{3m^{2} } \int d\omega
\sum_n \langle {\bf r}|\hat{{\bf p}} e^{i{\bf k\hat{r}}}|n\rangle\langle n|
\hat{{\bf p}}e^{-i{\bf k\hat{r}}}|{\bf r}'\rangle
\omega\delta(E - E_n -\omega).
\end{equation}
Because of the energy conservation the sum here includes only states 
$|n\rangle$ with energy $E_n$ below $E$. Consider for example
the tunneling from the ground $s$-state.
A dipole transition transfers the particle from the $s$-state to a $p$-state.
However, there are no quasidiscrete $p$-states $|n\rangle$ below the 
ground state in the potential well.
Scattering $p$-waves can penetrate the potential barrier from the continuum
with an exponentially small amplitude. This means that
$\Gamma({\bf r},{\bf r}';E)$ 
is again exponentially small if one or both arguments ${\bf r}$ and ${\bf r}'$ 
are under the barrier or 
inside the potential well. Therefore,  $\Gamma({\bf r},{\bf r}';E)$ does not 
considerably influence the tunneling amplitude.
The reason for that can be easily understood. A real radiation 
would be impossible
if there were no tunneling. Whence, the radiation width must vanish together
with the tunneling width. On the contrary, the real part of $\Sigma$ under
the barrier would present even if the tunneling probability would vanish. 

  To avoid misunderstanding we need to stress that the contribution to the
radiation intensity 
from the barrier area and the potential well, which was, in application to the
nuclear alpha-decay, the subject of recent
experimental \cite{exp} and theoretical \cite{Dyakonov,Bertch,B} 
studies, still may be important.
The radiation amplitude with $E_{s}-E_{p}=\omega$ contains the matrix element
\begin{equation}
\label{v}
<s|\hat{{\bf p}}|p>=\frac{1}{\omega}<s|[\hat{H},\hat{{\bf p}}]|p>=
\frac{i}{\omega}<s|\nabla \hat{U}|p>.
\end{equation}
When one moves inside the barrier from the outer turning point inwards, 
the resonance $s$-wave function exponentially increases while the non-resonance 
$p$-wave function exponentially decreases. As a result, the product 
$\psi_s(r)\psi_p(r)$ does not change considerably. This means that 
the contribution to the real radiation from the inner area may be
 comparable to that from the area outside the barrier. The gradient
$\nabla U$ changes its sign near the maximum of the potential which implies a
destructive interference between the radiation from the different 
areas (since $|s\rangle$ is the non-oscillating 
ground state wave function, the product $\psi_s(r)\psi_p(r)$ does not change
sign inside the barrier).  

This work was supported by the NSF grant 96-05207. V.F. gratefully acknowledges
the hospitality and support from the NSCL.

\end{document}